\documentclass[twocolumn,trackchanges]{aastex7}

\usepackage{amsmath, amsthm, amssymb,amsfonts}
\usepackage{graphicx}
\usepackage{dcolumn}
\usepackage{bm}
\usepackage{color}
\usepackage[abs]{overpic}
\usepackage{subcaption}
\usepackage{import}
\usepackage{longtable}

\begin{document}

\title{Evolution of an Alfvén Wave–Driven Proton Beam in the Expanding Solar Wind}

\author[gname=Jarrod,sname=Bianco]{J. S. Bianco}
\affiliation{Department of Physics, University of Texas at Austin, TX 78712, USA}
\affiliation{Institute for Fusion Studies}
\email{jarrodbianco@utexas.edu}

\author[gname=Anna,sname=Tenerani]{A. Tenerani}
\affiliation{Department of Physics, University of Texas at Austin, TX 78712, USA}
\affiliation{Institute for Fusion Studies}
\email{anna.tenerani@austin.utexas.edu}

 \author[sname=Carlos,gname=Gonzalez]{C. Gonzalez}
 \affiliation{Department of Physics, University of Texas at Austin, TX 78712, USA}
 \affiliation{Institute for Fusion Studies}
\email{carlos.gonzalez1@austin.utexas.edu}
  
\author[sname=Lorenzo,gname=Matteini]{L. Matteini}
  \affiliation{Department of Physics, Imperial College London, London SW7 2BW, UK}
\email{l.matteini@imperial.ac.uk}  

  \author[sname=Kristopher,gname=Klein]{K. G. Klein}
  \affiliation{Lunar and Planetary Laboratory and Department of Planetary Sciences, University of Arizona Tucson, AZ 85719, USA}
\email{kgklein@arizona.edu}

\begin{abstract}
We investigate the self-consistent formation and long-term evolution of proton beams in the expanding solar wind using an ensemble of one-dimensional hybrid expanding box simulations.  
Initial conditions are chosen to represent a range of plasma states observed by the Helios spacecraft at 0.3 AU, including an amplitude-modulated Alfv\'en wave that nonlinearly drives a proton beam aligned with the magnetic field. 
We compare simulation results with solar wind data out to 1.5 AU and show that our model reproduces key observed features of proton beams on average, such as the radial evolution of the drift and the relative core-to-beam density ratio. 
These findings support the theory that the observed evolution of the proton beam drift in the solar wind is determined by kinetic instabilities. 
More broadly, our results indicate that the interplay between nonlinear Alfv\'en wave dynamics, expansion effects and kinetic instabilities plays a fundamental role in solar wind dynamics, with implications for interpreting solar wind heating rate estimates.
\end{abstract}

\section{Introduction}
Proton beams are a ubiquitous feature of the Alfv\'enic solar wind. 
These beams drift along the mean magnetic field  at speeds near the local Alfv\'en speed, with respect to the core proton population, and are observed from regions close to the sun all the way out beyond 1~AU \citep{Marsch, goldstein2000observed, alterman2018comparison}. 
Characterizing their origin, evolution, and stability, and how they interact with waves, is essential for advancing our understanding of solar wind dynamics and heating.

Recent Parker Solar Probe observations reveal that proton beams are often found at radial distances less than 0.2 AU and near the heliospheric current sheet \citep{verniero2020parker,ofman2022modeling,phan2022parker,Ofman_2025}, not necessarily within Alfv\'enic streams. However, they display distinct characteristics that vary with the type of turbulent environment: in Alfv\'enic (high cross-helicity) intervals, beams tend to be cooler than the core and exhibit reduced velocity-space scattering compared to those in non-Alfv\'enic (low cross-helicity) streams \citep{verniero2022strong, gonzalez2024local}. 
These findings suggest that different turbulence regimes drive different wave-particle interactions and kinetic signatures. 
In particular, proton beams in Alfv\'enic wind may result from the coupling of Alfv\'enic fluctuations with compressible modes and the formation of steepened wave fronts, which are often observed as arc-polarized structures in these streams \citep{tsurutani2005nonlinear}. 
In this regard, numerical simulations have shown that field-aligned proton beams can arise from nonlinear compressible processes involving Alfv\'en waves, such as Alfv\'en wave steepening and collapse or modulational instabilities \citep{spangler_nonlinear_1985, machida_simulation_1987, velli1999propagation, buti2000hybrid, gonzalez2021proton, gonzalez2023particle}, and parametric  instabilities \citep{araneda2008proton, matteini2010kinetics, maneva2013turbulent, maneva2015relative}. 

Departures of the particle velocity distribution functions (VDFs) from Maxwellianity can trigger phase-space instabilities. 
In the solar wind, adiabatic expansion drives VDFs towards both firehose and proton core-beam drift instabilities, a prediction that is also supported by numerical simulations \citep{matteini2006parallel,hellinger2008oblique, hellinger2011proton, hellinger2019turbulence,Ofman_2014,ofman2017effects,ofman2022modeling,maneva2013turbulent,maneva2015relative}. In-situ spacecraft measurements indicate that proton VDFs evolve with radial distance, however, the observed evolution deviates from predictions based solely on adiabatic expansion \citep{matteini_signatures_2013}. 
Moreover, proton distribution moments---including temperature anisotropy, plasma beta, and relative core-beam drift and density---appear to be limited by instability thresholds \citep{hellinger_solar_2006, bale2009magnetic, goldstein2000observed, tu_dependence_2004, huang2020proton,martinovic2021ion}, {indicating that not only heating mediated by the turbulent cascade (e.g. through stochastic heating, wave-particle resonances, or reconnection at intermittent structures \citep{Chandran_2010,bowen_mediation_2024,Osman_2014}), but also wave-particle interactions triggered by kinetic instabilities play a key role in regulating solar wind properties and energetics.}

{Previous studies using the hybrid expanding box model have examined expansion-driven instabilities in a single proton population \citep{matteini2006parallel,hellinger2008oblique}, including cases with turbulent fluctuations \citep{2017Hellinger,hellinger2019turbulence}, as well as in prescribed core-beam systems \citep{hellinger2011proton}. 

Maneva et al. \citep{maneva2013turbulent,maneva2014regulation,maneva2015relative} performed both 1D and 2D simulations of multi-ion plasmas initialized with broadband or monochromatic Alfvén wave spectra. In particular, \citet{maneva2013turbulent,maneva2014regulation} investigated the heating of minor ions (alpha particles) and the regulation of relative proton–alpha drifts via parametrically unstable Alfvén waves in 1D. The self-consistent development of a proton beam has been reported in 2D simulations of a proton-alpha plasma, attributed to parametric instabilities \citep{maneva2013turbulent}; these studies reported that, when the initial spectrum is turbulent, solar wind expansion tends to slow down the relative drift between ion species \citep{maneva2015relative}. While these studies address differential ion streaming and instability-driven relaxation, they do not systematically examine proton beam formation and its long-term stability under expansion. 

Despite previous theoretical and observational efforts, a comprehensive study of the self-consistent generation and evolution of proton beams under the combined influence of kinetic instabilities and solar wind expansion—together with a direct comparison to spacecraft observations—remains missing.}

To address this gap, in this work we make use of a set of one-dimensional hybrid particle-in-cell simulations that incorporate expansion effects through the expanding box model. 
We select initial conditions that are representative of a wide range of plasma states observed by the Helios spacecraft at 0.3 AU and compare simulation results with data out to about 1.5 AU. 
We show that, despite the simplified geometry, our model reproduces on average the main observed characteristics of proton beams such as the relative beam to core density and the evolution of normalized drift speed. 
We also discuss the evolution of adiabatic invariants by proposing a possible interpretation for parallel heating rate estimates.

In the following, we describe the model and methods in section \ref{methods}; numerical results including linear stability analysis are reported in section \ref{results}; we compare numerical outputs with solar wind data in section \ref{data_comparison} and provide a final summary and discussion in section \ref{discussion}. 
 
\section{Numerical Model and Methods}
\label{methods}

To study the formation and evolution of proton beams in the solar wind, we make use of the expanding box model \citep{velli1992mhd,grappin1996waves,liewer2001alfven} implemented in the hybrid particle-in-cell code CAMELIA \citep{hellinger2003hybrid,franci_three-dimensional_2018}, which is based on the CAM-CL algorithm \citep{matthews_current_1994}. 
In this model, curvature effects are neglected but the main effects of radial expansion are retained through an expanding metric in the transverse plane $(y,z)$ and source terms in the model equations that arise when moving to the wind co-moving frame \citep{liewer2001alfven}. 
By assuming a supersonic and super-Alfv\'enic wind with constant radial speed $U_0$, the solar wind expansion enters plasma dynamics through the scaling factor 
\begin{equation}
    a(t) = \frac{R(t)}{R_0} = 1 + \frac{U_0}{R_0}t = 1 + \frac{t}{t_{exp}},
    \label{eqn:a_t}
\end{equation}
where $R$ represents the mean radial distance of the plasma parcel (the simulation box) from the Sun, $R_0$ is the initial radial distance, and $t_{exp}=R_0/U_0$ is the expansion time.
The code employs periodic boundary conditions.
Lengths are normalized to the initial ion inertial length, $d_i$ $=$ $c/\omega_{pi}$ with $\omega_{pi}$ $=$ $\sqrt{4\pi ne^2/m_i}$, time is normalized to the initial inverse proton gyrofrequency, $\Omega_{ci}^{-1}$ $=$ $(eB_0/cm_i)^{-1}$, and velocity to the initial Alfv\'en speed, $v_a$ $=$ $B_0/\sqrt{4\pi nm_i}$. 

We perform five one-dimensional simulations of a left-handed Alfv\'en wave propagating along the mean magnetic field ${\bf B}_0$ that is taken along the $\hat x$-axis (radial direction). 
The plasma is otherwise locally uniform, with an initial bi-maxwellian distribution for protons and isothermal electrons. 
The initial proton parallel plasma beta $\beta_{\parallel}=8\pi nT_{\parallel}/B_0^2$, proton temperature anisotropy $T_\perp/T_\parallel$ and numerical parameters are reported in Table \ref{tab:collapse_param} for each simulation. {Although in the expanding box model the total energy is not conserved, energy evolves following a generalized conservation law of the form $dE/dt = f(R)$, where $f(R)$ is the energy decay rate introduced by expansion effects on magnetic and mass flux, and particles momentum. We verified that for our chosen numerical parameters, the total energy evolves as expected with a maximum absolute error $|dE/dt - f(R)|\approx 10^{-5}$ across all the simulations, dropping below $10^{-6}$ after $t\gtrsim5000\Omega_{ci}$.} 
The plasma beta and temperature anisotropy are based on values observed in the Alfv\'enic wind at $R\approx0.3$ AU \citep{vdurovcova2019evolution} and are chosen to represent a range of solar wind conditions with the exception of run RC, which was initialized with a high initial temperature anisotropy. 
RC is of interest to test the effects of the ion-cyclotron instability on the evolution of the system with a growth rate higher than $\gamma/\Omega_{ci}=10^{-3}$, as this instability can be triggered within the turbulent cascade if a sufficiently high $T_\perp/T_\parallel$ is achieved \citep{zhang2025extreme}. 
The electron beta is initially set equal to $\beta_\parallel$. 

The wave magnetic field is given by
\begin{equation}
    B_y(x)=A(x)\cos(kx),\quad B_z(x)=A(x)\sin(kx),
\end{equation}
with the corresponding bulk velocity fluctuations determined by the ion-cyclotron wave dispersion relation. 
We impose a Gaussian-modulated amplitude of the form $A(x) = \delta b\; \exp\left[-\left({x}/{l}\right)^2\right]$, where $l$ is the modulation length scale, chosen to be $l=0.2 L_x$ for all runs, and the initial wave amplitude being $\delta b=0.5B_{0}$. 
The  main wave number is $n=10$ and thus $k d_i = 0.25$. 
Although Alfv\'enic wind displays fluctuations characterized by a nearly constant magnetic field strength \citep{barnes1974large}, we choose an amplitude-modulated wave packet to clearly isolate the process of beam formation through wave-steepening from other processes that might otherwise take place concurrently, like the parametric instability (e.g., \citet{gonzalez2023particle}). 
As explained later, the particular form of the Alfv\'en wave does not affect the formation of the beam.

\begin{table}[h]
    \centering
    \begin{tabular}{cccccc}
    \hline
      Runs & $\beta_{\parallel}$ & $T_{\perp}/T_{\parallel}$ & $N_x$ & $\Delta_x$ & $\Delta_t$\\
       RB  & 0.25 & 1.75 & 1024 & $d_i$/4 & 0.025 $\Omega_{ci}^{-1}$\\
       R2  & 0.25 & 1 & 1024 & $d_i$/4 & 0.050 $\Omega_{ci}^{-1}$\\
       R3  & 0.1 & 2 & 2048 & $d_i$/8 & 0.025 $\Omega_{ci}^{-1}$\\
       R4  & 0.4 & 2 & 1024 & $d_i$/4 & 0.020 $\Omega_{ci}^{-1}$\\
       RC  & 0.05 & 4 & 2048 & $d_i$/8 & 0.025 $\Omega_{ci}^{-1}$\\
    \hline
    \end{tabular}
    \caption{Initialization parameters: parallel plasma beta, temperature anisotropy, number 
    of cells, cell size, and particle time step respectively. 
    The fields are advanced every 10 substeps and each run contains $N = 10^4$ particles per cell. 
    All runs will achieve a maximum run time of $t=30000 \Omega_{ci}^{-1}$.}
    \label{tab:collapse_param}
\end{table}

The expansion rate $\dot a=1/t_{exp}$ is set to $\dot{a} = 10^{-4} \Omega_{ci}$ 
for all runs, ensuring a sufficiently large separation of scales between the expansion rate and the wave frequency consistent with solar wind parameters. Comparing to the estimated expansion rate for fast solar wind ($v_{sw}\approx700\;km/s$) at 0.3 AU gives $\dot{a}_{sw} = 6\times10^{-5}\Omega_{ci}$ (using the value of the proton cyclotron frequency, $f_{ci} = 0.615\;s^{-1}$ from Helios data at 0.3 AU) which is slightly lower than the simulation rate but within the same order of magnitude.

To study the growth of kinetic instabilities throughout the course of each simulation, we use the New Hampshire Linear Dispersion Relation solver (NHDS; \cite{Verscharen_2018}). 
Assuming a bi-maxwellian distribution function for each input particle species,  
the code solves the hot-plasma dispersion relation for real and imaginary components of the frequency $\omega = \omega_r + i\gamma$ where $\gamma$ is the growth rate for unstable modes in units of $\Omega_{ci}$. 
Because in our system the proton beam forms dynamically and evolves with expansion, the single bi-maxwellian description might not provide the best representation of the proton VDF. 
Therefore, to buffer times within the simulations where NHDS does not provide reliable results, we complement our analysis with the Arbitrary Linear Plasma Solver (ALPS; \cite{verscharen_alps_2018}). 
ALPS similarly solves the hot plasma dispersion relation but for an arbitrary input VDF. 
In this case, we take the output VDF from selected times where we expect a transition from core-beam towards temperature anisotropy-driven instabilities as the input of ALPS. The most unstable modes in those cases are searched using the output map of the dielectric function in frequency space. A parallel wave vector scan is then performed over the minima to investigate growth rates.  

Finally, to analyze the core-beam populations in our simulations, we manually split the system into two distinct species, when a beam can be identified as a distinct population. 
To this end, we find the drift speed of the beam with respect to the core by inspecting the VDF in the parallel velocity space $v_\parallel$, and then extract the beam as an approximate Gaussian distribution. The core population is obtained by subtracting the beam distribution from the total VDF. 
We also make use of the identified range of values in $v_\parallel$ that belong to the beam to extrapolate the beam also in the $v_\perp$ direction. 
This allows us to calculate the core and beam velocity moments via numerical integration with respect to both $v_\parallel$ and $v_\perp$, a common approach to analyze solar wind VDFs \citep{klein2021inferred}. 

Core and proton beam parameters obtained with this method are then compared with Helios and Ulysses data. For Ulysses, we use fast wind ($v_{sw}$ $>$ $650$ $km/s$) data taken by the SWOOPS investigation during two out-of-ecliptic passes, from 1994-1995 and 2006-2007, respectively, at around 1.5 AU radial distance. 
The Ulysses dataset contains proton VDF moments and bi-Maxwellian fit parameters, including thermal velocities, core and beam proton densities and temperature anisotropies, and bulk velocities, at 4 minute resolution. 
For Helios, we select fast wind data in the range of radial distances $R=0.3 -1$ AU. 
This dataset consists also of bi-Maxwellian fit parameters of the proton VDFs, measured by the E1 experiment onboard Helios 1 and Helios 2. 
The dataset includes proton core and beam density, bulk velocity and temperatures at $\approx 40.5$ second resolution \citep{vdurovcova2019evolution}. 

\section{Simulation Results}
\label{results}

All simulations in this work have a common initial stage dominated by transient compressible effects in which a proton beam and ion acoustic modes are generated, followed by a slow, expansion-driven evolution, where kinetic instabilities regulate wave-particle interactions. 
We describe the two evolutionary stages in sections \ref{stage1} and \ref{stage2}, respectively, followed by an analysis of kinetic instabilities and discussion on their role in regulating VDF properties in section \ref{alps}.

\subsection{Formation of Proton Beams and Ion-Acoustic Modes}
\label{stage1}

\begin{figure*}[h]
\centering
\includegraphics[width=0.8\linewidth]{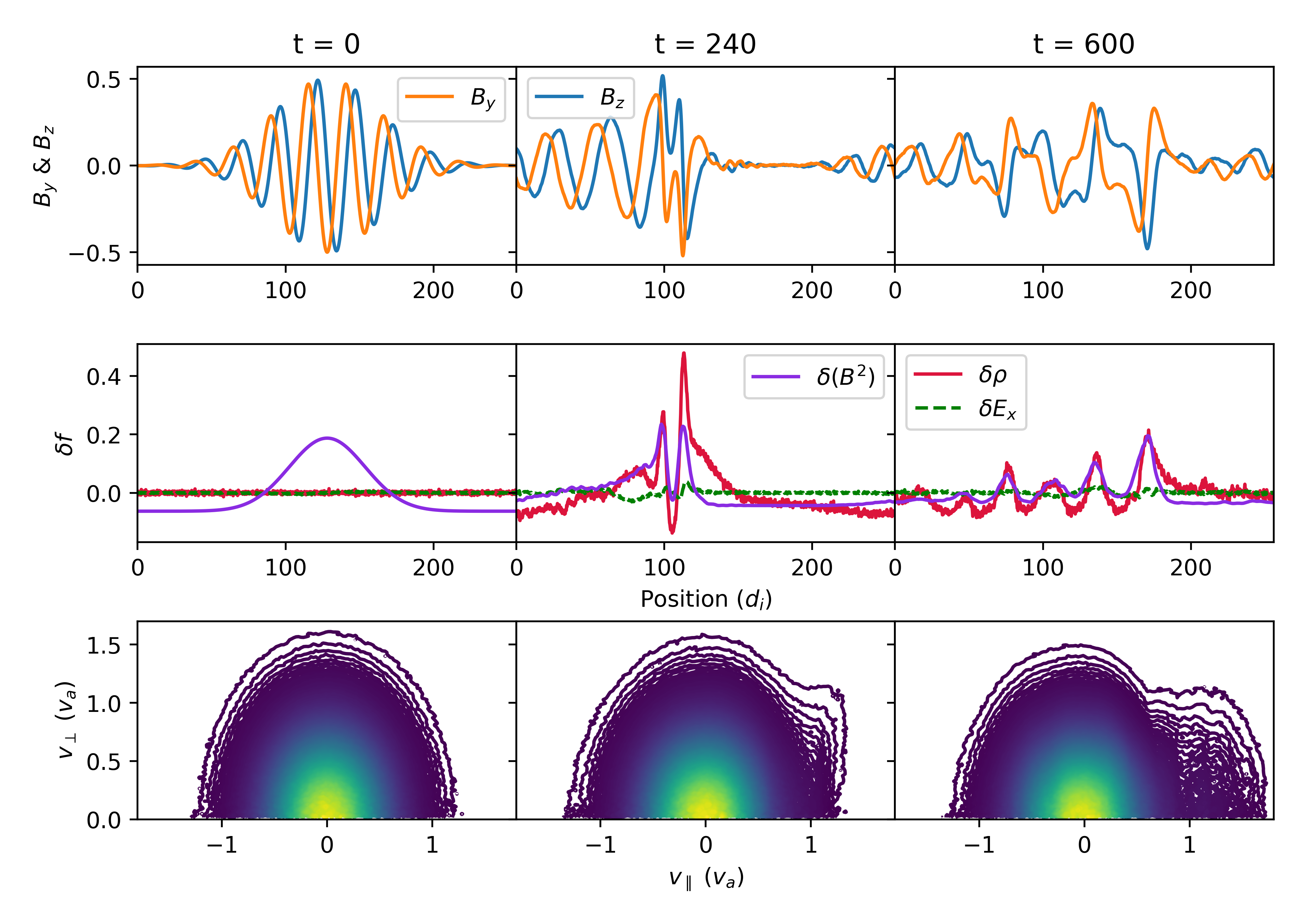}
\caption{Snapshots taken from simulation RB displaying the Alfvén wave collapse and proton beam formation. Top row: magnetic field components $B_y(x)$ and $B_z(x)$; middle row: fluctuations of the magnetic field magnitude $B^2(x)$, field-aligned electric field $E_x(x)$, and density $\rho(x)$; bottom row: contour plot of the VDF in the $(v_\perp,v_\parallel)$ space.}
\label{fig:collapse}
\end{figure*}
\begin{figure*}[h]
\centering
\includegraphics[width=1\linewidth]{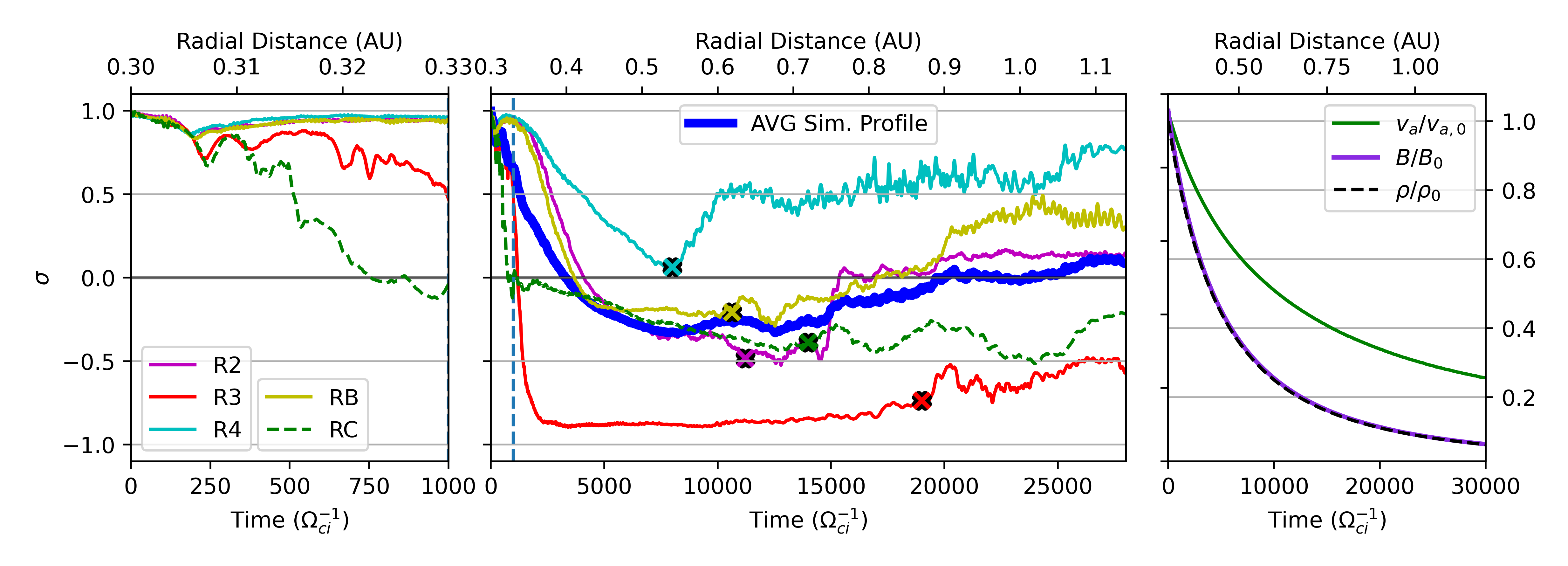}
\caption{Normalized cross-helicity $\sigma$ as a function of time and radial distance for all simulations. 
The left panel shows a close-up view of $\sigma$ at early times. 
The blue profile in the right panel shows $\sigma$ averaged over all runs. The vertical blue dashed line indicates time $t=1000\;\Omega_{ci}^{-1}$ as a reference marker for the left panel. The x's represent the times just before each run became firehose unstable. The right panel shows $v_a$, B, and $\rho$ averaged simulation profiles as a function of time and radial distance.}
\label{fig:crosshl1}
\end{figure*}

The initial evolution of the wave packet is characterized by nonlinear wave steepening that causes the wave-packet to collapse on itself. Because of the spatially-dependent amplitude, the central part of the wave propagates faster than the leading and trailing edges. As the wave ``runs into itself", steepening occurs on the leading edge of the wave where small scales are formed. In turn, the steepened edge of the wave further slows down due to the dispersive nature of left-handed waves ultimately leading to the wave collapse. Although here we have considered an amplitude-modulated wave-packet, the steepening and collapse process is quite general for left-handed, weakly dispersive waves. It has been described theoretically through the Derivative Nonlinear Schroedinger equation \citep{spangler_nonlinear_1985}, investigated with hybrid simulations \citep{velli1999propagation,buti2000hybrid, matteini2010kinetics}, and it was also observed in simulations of both broadband and monochromatic Alfv\'en waves with initial constant magnetic field strength \citep{gonzalez2021proton,gonzalez2023particle}.  

This process is displayed in Figure~\ref{fig:collapse}, where in the top row we show three snapshots of the wave magnetic field as a function of $x$. 
As can be seen, at around $t=240$ (middle top panel), the wave collapses distributing the wave power to larger wave numbers (spectrum not shown here). 
The wave collapse leads to the formation of forced compressible perturbations of the fast type, with sharp density compressions in phase with fluctuations of the magnetic pressure, shown in the second row in red and purple color, respectively. 

The compressible structures are associated with a field-aligned electric field (green color) propagating at nearly $v\simeq 0.87 v_a$. This was found by tracking the slope of the wave front in $x-t$ space. 
The propagating electric field structure, by acting as a snowplow,  accelerates protons away from the center of the particle velocity distribution function into a field-aligned beam drifting at nearly the local Alfv\'en speed \citep{machida_simulation_1987, matteini2010kinetics}. 
The formation of the beam can be seen in the last row of Figure~\ref{fig:collapse}, where the contours of the proton VDF in the $(v_\parallel,v_\perp)$ plane are plotted.



With our setup, the formation of the beam takes place on a short time scale, between $t\approx 100-200$, with the collapse occurring somewhat faster for the larger initial $\beta_\parallel$ (RB, R2 and R4). 
After the formation of the proton beam at a speed $v_d\gtrsim v_a$, the beam slows down approaching the local Alfv\'en speed (see also Fig.~\ref{fig:temp_drift}, left panel). 
The role of kinetic instabilities in regulating the core-beam drift speed will be addressed in the next section.   

In addition to the steepening and subsequent collapse, the wave packet is subject to parametric decay. 
Parametric decay is the instability of a large amplitude Alfv\'en wave in which the wave resonates with, and transfers energy to, a reflected Alfv\'en wave and a forward propagating ion-acoustic mode \citep{Derby1978,jayanti1993dispersion}. 
Parametric decay occurs on timescales longer than the collapse, and becomes slower and weaker for larger $\beta_\parallel$, as predicted from theory and simulations \citep{del2001parametric,gonzalez2020role}. 
For reference, in Figure \ref{fig:crosshl1} we report the  normalized cross-helicity $\sigma$ as a function of radial distance and time,

\begin{equation}
    \sigma = \frac{<|\delta{\bf z}^-|^2>_x - <|\delta{\bf z}^+|^2>_x}{<|\delta{\bf z}^+|^2>_x + <|\delta{\bf z}^-|^2>_x},
    \label{eqn:crosshel}
\end{equation}
where $\delta {\bf z}^\pm$ are the Elsasser variables and $<\cdot>_x$ represents spatial average. 
Wave reflection, one of the signatures of parametric instabilities, is observed for times up to $t=5000$ where for the cases of lower initial parallel beta (R3), $\sigma$ drops to around $\sigma=-1$, indicating complete reflection, whereas for higher initial parallel beta (R4) reflection is slower and weaker. 
On average over all runs, the cross-helicity floats just below $\sigma=0$.   

The growing ion-acoustic waves formed through the parametric decay eventually trap particles at approximately the sound speed, $c_s = v_a\sqrt{(\beta_e + \beta_i)/2}$ thereby saturating the instability \citep{matteini2010kinetics}.

\begin{figure}[h]
\includegraphics[width=\linewidth]{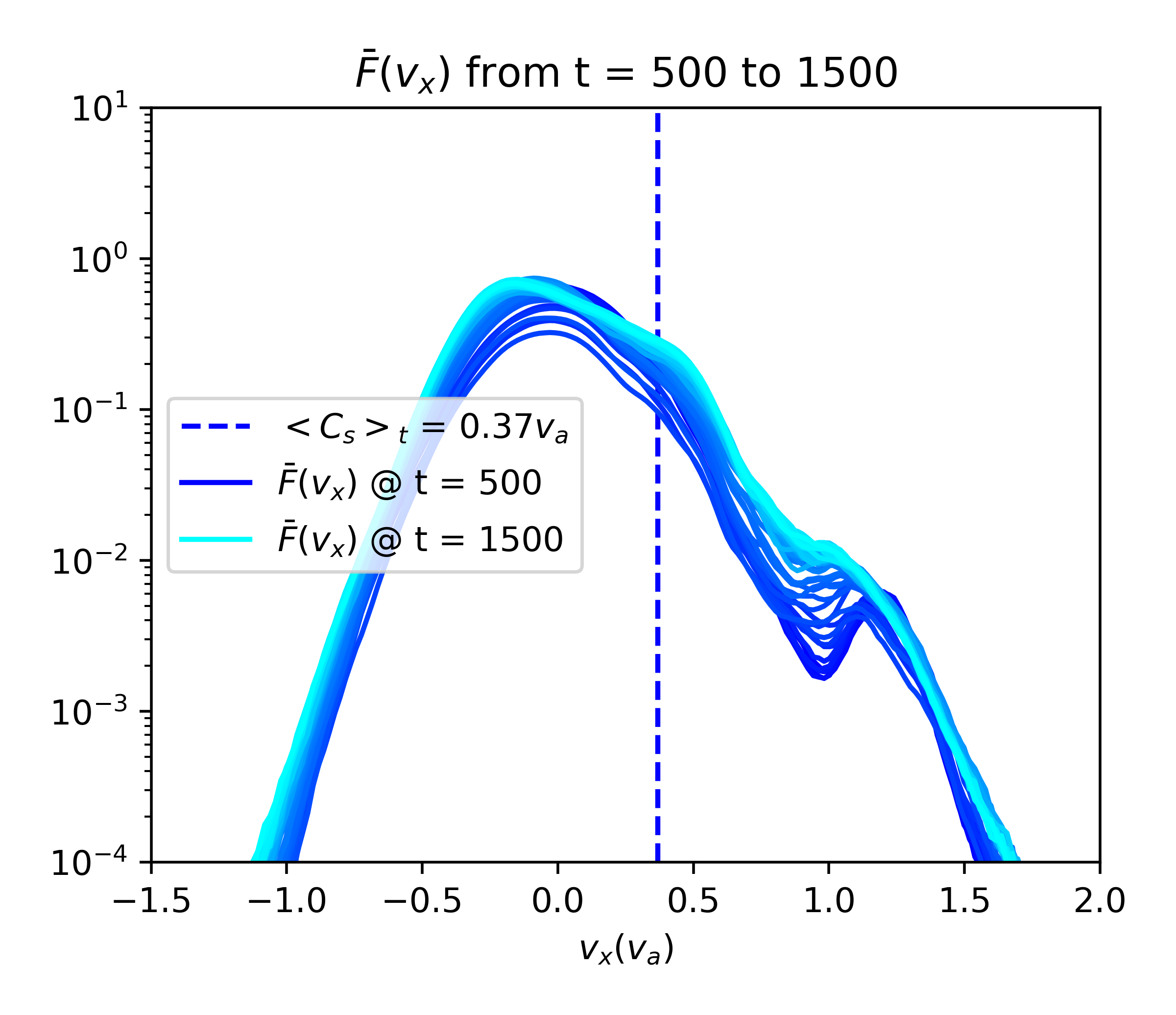}
\caption{The time evolution of the distribution function  $F(v_x)$ from $t=0$ to $t=1500$ for R3, showing the formation of a trapped population and a beam that slows down to the local Alfv\'en speed.}
\label{fig:vdftrapping}
\end{figure}

Signatures in the proton VDF of the two mechanisms discussed in this section, i.e., wave steepening/collapse and parametric decay, can be seen in Figure~\ref{fig:vdftrapping}. 
Here we plot the proton VDF as a function of $v_x$ at different times until $t=1500$, showing that the proton beam slows down to $v_x\approx v_a$ and that a second population of trapped particles forms at $v_{x} \approx <c_s> =0.37 v_a$. Here $<c_s>$ is the average sound speed calculated from averaging the values of $c_s$ from times 500 to 1500. 



\subsection{Expansion Driven Evolution}
\label{stage2}

\begin{figure*}[t]
\centering
\includegraphics[width=\linewidth]{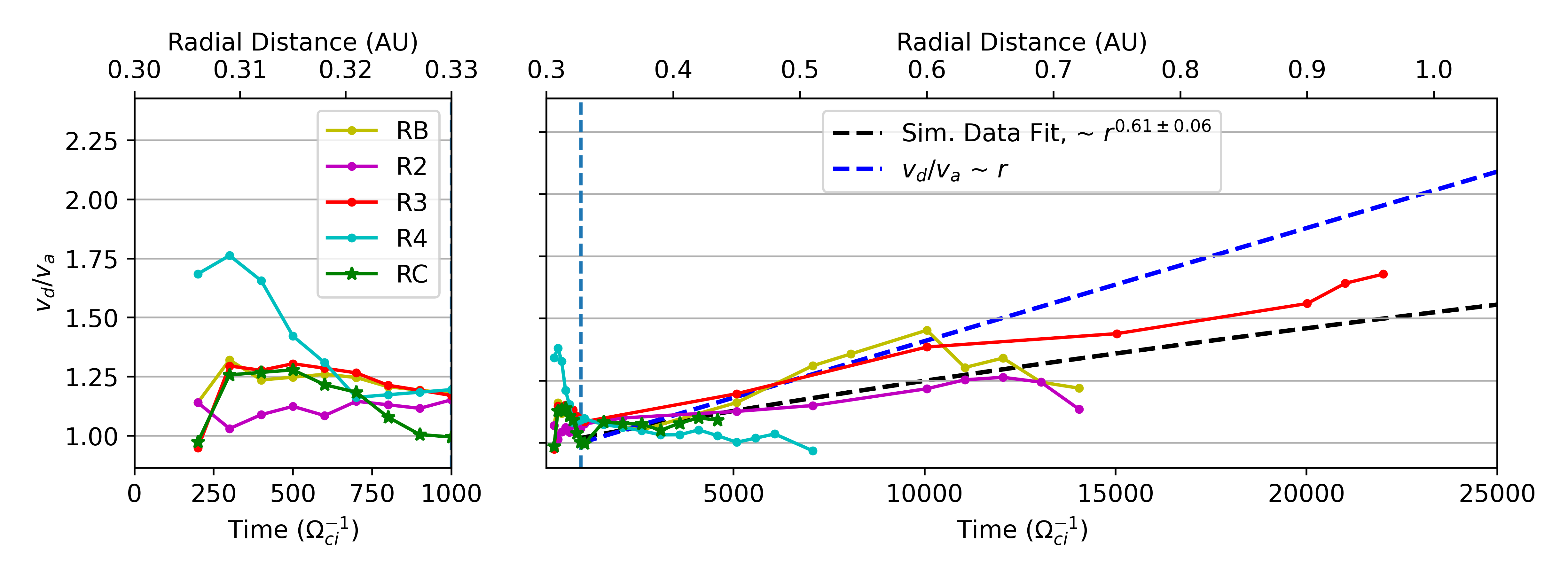}
\caption{The normalized drift speed $v_d/v_a$ for each run as a function of time and radial distance.  
The vertical blue line in the right panel indicates $t$ = 1000 $\Omega_{ci}^{-1}$ which marks the end of the initial transient stage. 
The black line shows the best fit to the whole ensemble of data points and the dark blue line represents the CGL profile. 
On the left is a zoom into early times.}
\label{fig:temp_drift}
\end{figure*}

As the simulations continue to evolve, a quasi-stable state is reached, where the drifting proton beam and core population coexist, while the system slowly evolves due to the interplay of kinetic instabilities and expansion. 
In fact, expansion concurrently drives the evolution of both the normalized core-beam drift and the temperature anisotropy \citep{hellinger2011proton}. 

With a radial mean magnetic field as considered in this work, the Alfv\'en speed decreases linearly with radial distance, $v_a\propto(R/R_0)^{-1}$ (Fig.~\ref{fig:crosshl1}, right panel), while the proton's radial momentum is conserved. 
Therefore, the normalized beam drift increases with radial distance, $v_d/v_a\propto(R/R_0)$, becoming a source of unstable waves as the beam drift crosses instability thresholds. In Figure~\ref{fig:temp_drift} we show the extrapolated drift of the beam with respect to the core, normalized to the local Alfv\'en speed $v_a$, as a function of time and radial distance. 
The left panel shows a close-up view of $v_d/v_a$ at early times when the beam initially forms and rapidly slows down. 

After the system has adjusted to a quasi-stable state, it enters the expansion-driven stage at about $t=1000$, marked by the vertical blue dashed line in the right panel. 
We find that during this stage the evolution of $v_d/v_a$ depends on the specific plasma state. In this expansion-driven phase, runs RB and R3 display a nearly linear radial increase of the normalized drift until $t\approx10000$, denoted by the solid blue line $v_d/v_a\propto R$. 
However, as an ensemble over the simulations, the normalized drift increases slower than linear, and it is best fit by $v_d/v_a\propto(R/R_0)^{0.61\pm0.06}$. This is likely due to kinetic instabilities enabled by the core-beam system that slow down the relative drift.
A quantitative study of kinetic instabilities is reported in section \ref{alps} to support this interpretation.

\begin{figure}[h]
\centering
\includegraphics[width=1\linewidth]{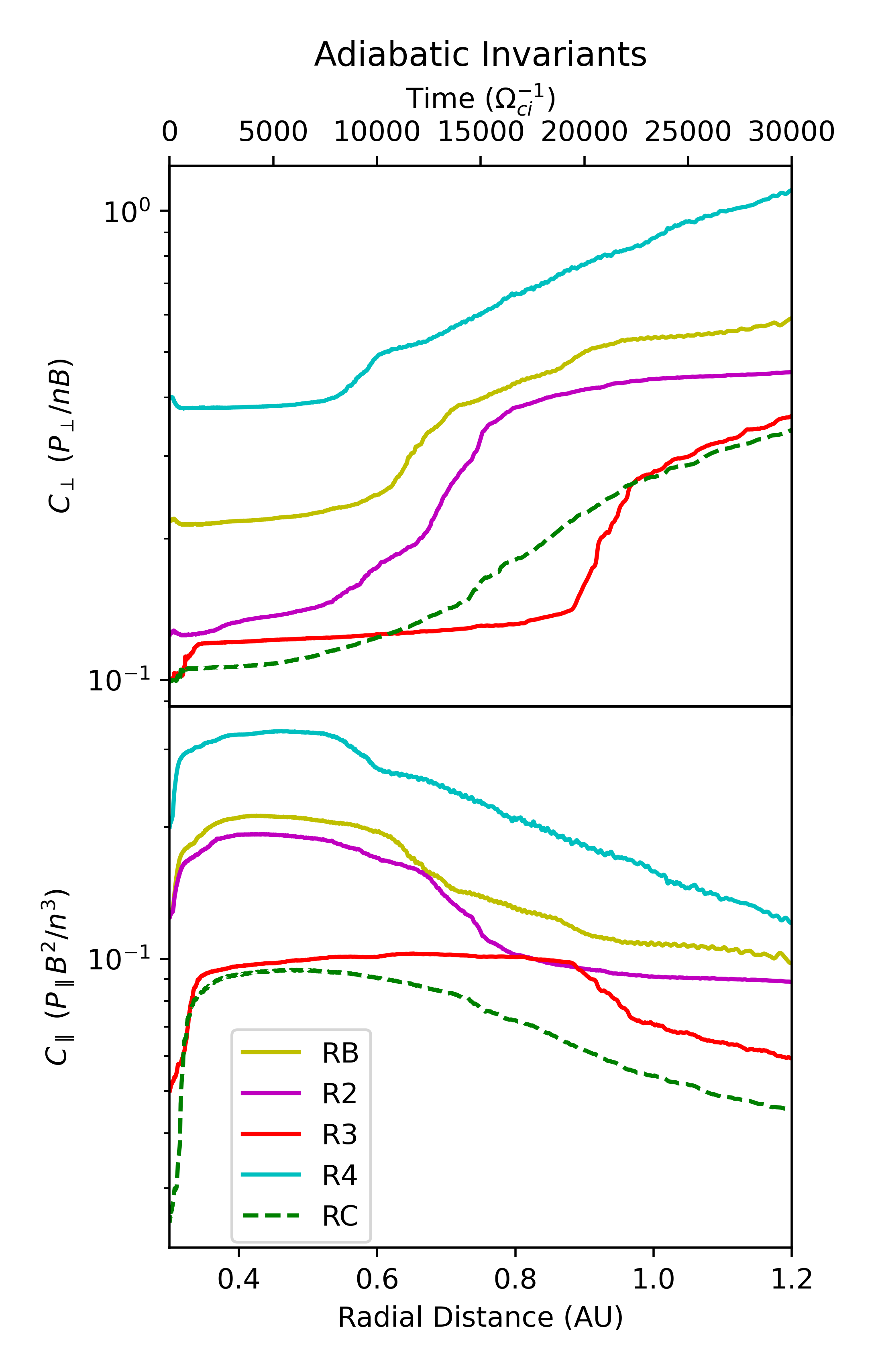}
\caption{The two adiabatic invariants (Eqn. \ref{eqn:cglpars}) calculated for each run.}
\label{fig:invariants}
\end{figure}

In the absence of heating and strong deviations from a bi-Maxwellian of the proton VDF, the thermal evolution of the plasma is expected to follow the CGL, double adiabatic prediction (\cite{chew_boltzmann_1956}) with two invariants,
\begin{equation}
C_{\perp} = \frac{P_{\perp}}{nB} \;\;\;\&\;\;\; C_{\parallel} = \frac{P_{\parallel}B^2}{n^3},
\label{eqn:cglpars}
\end{equation}
which are reported for the total VDF pressures and density in Figure \ref{fig:invariants} for all runs. 
After the initial transients discussed in the previous section, ending at  $t\approx 1000$, the runs hold the CGL prediction relatively well up until $t=7000-20000$, depending on the specific initial conditions (in practice, until the firehose instability onset, as discussed later). 
{On average, in the time interval $1000\lesssim t\lesssim 10000$, $C_\parallel$ is essentially constant and around $C_\parallel\approx 1.4\times 10^{6}\, KnT^{2}cm^{6}$ ($\approx 2\times 10^{-47}\, JT^2m^6$). This was found by converting the normalized code units using values of $B = 40\;nT$ and $n = 25\;cm^{-3}$ from Helios Solar Wind data at 0.3 AU.} 

Contrary to the beam drift speed, during this intermediate stage the total $\beta_\parallel$ follows to a good approximation the adiabatic prediction $\beta_\parallel\propto(R/R_0)^2$ (not shown here). 
Thus, in this case, the conservation of the two adiabatic invariants in a radially expanding medium implies that $T_\perp/T_\parallel\propto (R/R_0)^{-2}$, so that the system slowly evolves towards the firehose instability threshold (figure \ref{fig:parspace1}).

\begin{figure}[h]
\centering
\includegraphics[width=\linewidth]{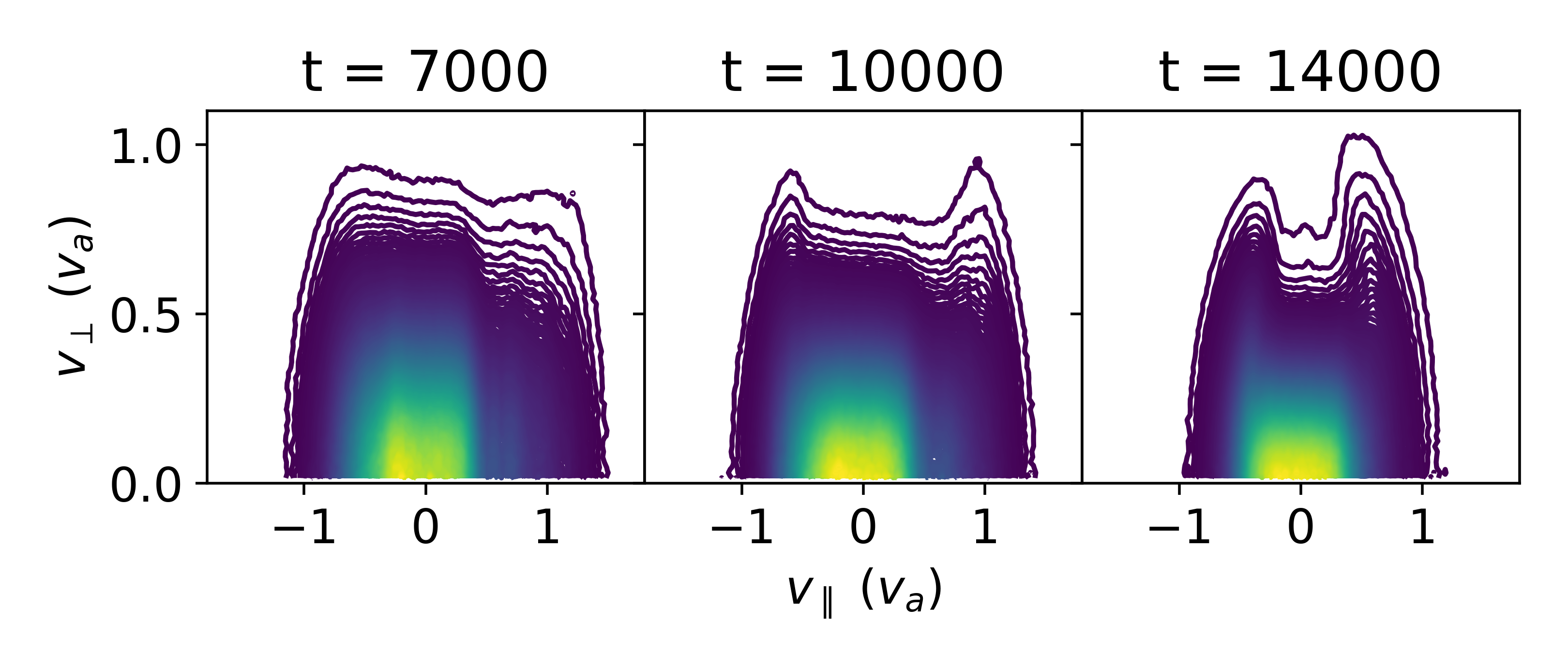}
\caption{Evolution of the VDF for RB, showing that the core-beam structure collapses due to the onset of the fire-hose instability.}
\label{fig:collapseVDF}
\end{figure}

The onset of the firehose instability leads to a reduction in the overall proton temperature anisotropy. 
This instability drives an increase in $C_\perp$ and a corresponding decrease in $C_\parallel$, as evident in each simulation run through inspection of Figure \ref{fig:invariants}. 
The development of the firehose is also marked by the evolution of the VDF from a core-beam structure to a single ``double-horned" or pinched shape (\cite{Matteini_2015,markovskii_effect_2022}). 
Figure \ref{fig:collapseVDF} illustrates this transition for run RB, although a similar evolution is observed across all runs. 
Thus, on the long term, the firehose instability driven by the total temperature anisotropy of protons supersedes the dynamics of the core-beam system and the beam is scattered in phase space. 

\subsection{Linear Analysis of Kinetic Instabilities}
\label{alps}

\begin{figure}[h]
\centering
\includegraphics[width=1\linewidth]{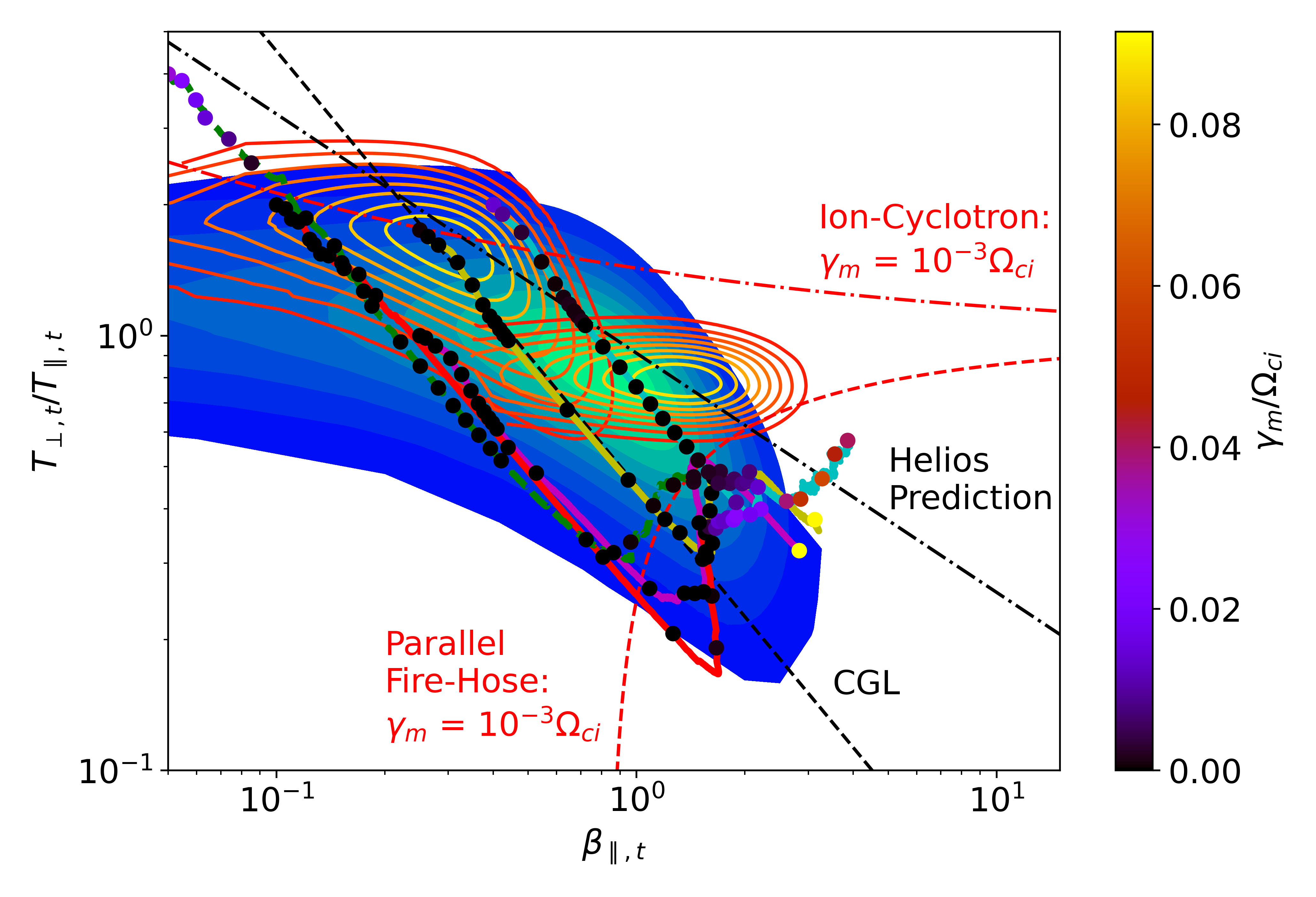}
\caption{Parameter space paths with the background taken from Ulysses (green contour) and Helios (red and blue for 0.3 and 1 AU respectively) fast wind data. 
The paths are color coded with dots taken from simulation data as the maximum growth rate at that point in time obtained from NHDS. 
Red lines are contour lines of $\gamma/\Omega_{ci}=10^{-3}$ for temperature anisotropy-driven instabilities. 
The dashed line indicates the CGL prediction and the dash-dotted line is the path followed by the centroid of the data distribution obtained by fitting Helios data as in \citep{matteini_signatures_2013}.}
\label{fig:parspace1}
\end{figure}

The stability of the total proton VDF depends on the  parallel plasma beta and temperature anisotropy $(\beta_{\parallel}, T_\perp/T_\parallel)$. 
In Figure \ref{fig:parspace1}, each run is plotted in that parameter space for selected times to help visualize the evolution of the total VDF.
In this plot, the contours represent Helios and Ulysses fast solar wind data with the green contours being Ulysses data at 1.5 AU, red being Helios data at 0.3 AU, and the blue solid contours being Helios data at 1 AU. 
Total parallel beta and anisotropy are calculated from in-situ measurements of proton temperature and magnetic field magnitude and are plotted on a grid, colored by the concentration of points relative to the maximum data concentration. 
The contours are then cutoff at 0.1 of the maximum. 
The points overlaid to the contours represent the path of each simulation through the parameter space, colored by maximum growth from NHDS. 

\begin{figure}[h]
\centering
\includegraphics[width=\linewidth]{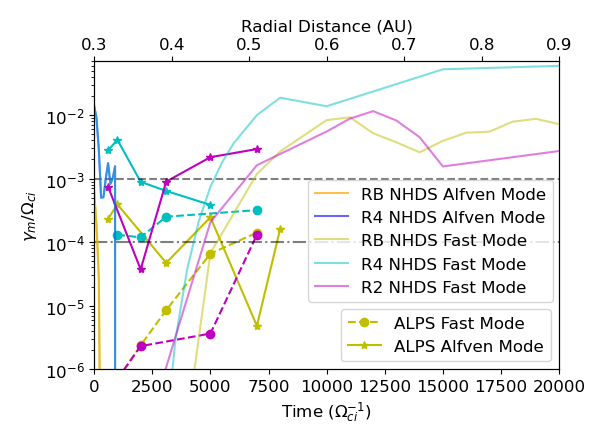}
\caption{The maximum growth rate, $\gamma_m/\Omega_{ci}$, for RB (yellow), R2 (magenta), and R4 (cyan) over time and radial distance. 
The solid lines are data points from NHDS and the dots and stars are from ALPS.}
\label{fig:growthrates}
\end{figure}

\begin{figure}[h]
\centering
\includegraphics[width=1\linewidth]{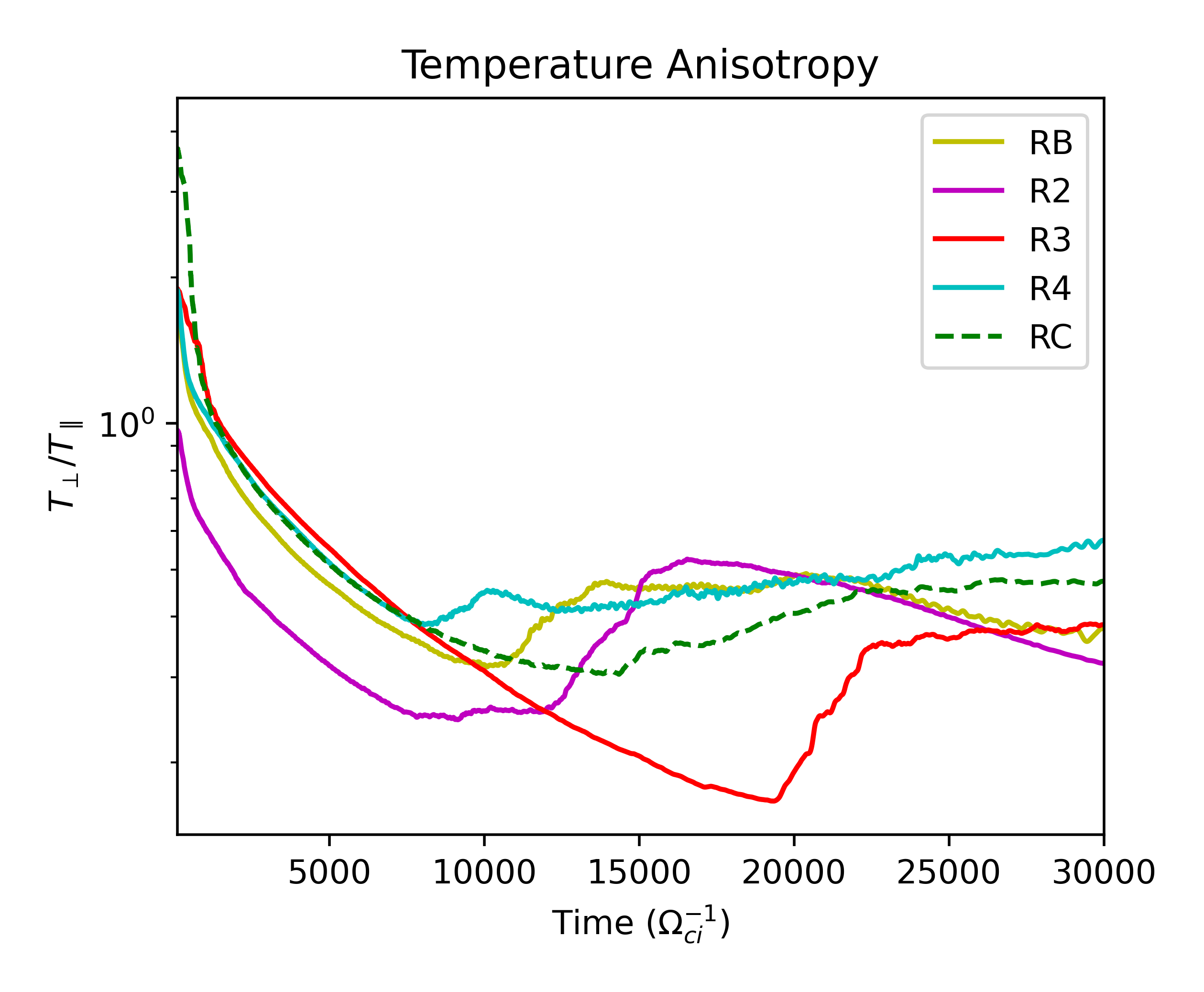}
\caption{Temperature anisotropy, $T_{\perp}/T_{\parallel}$, for each simulation plotted as a function of time.}
\label{fig:anisotropy}
\end{figure}

To complement Figure \ref{fig:parspace1}, Figure \ref{fig:growthrates} shows the  evolution of the maximum growth rate $\gamma_m$ obtained from NHDS (solid colored lines) and ALPS (lines with symbols) normalized to the local proton-gyrofrequency as a function of time (lower axis of abscissas) and radial distance (upper axis of abscissas). 
The dash-dotted line indicates  $\gamma_m/\Omega_{ci}=10^{-4}$, comparable to the expansion rate, and the dashed line is $\gamma_m/\Omega_{ci}$ = $10^{-3}$. 
The colors yellow, blue, and magenta represent runs RB, R4, and R2 respectively. 

By inspection of Figures \ref{fig:parspace1} and \ref{fig:growthrates}, it is apparent that according to the NHDS linear analysis the system is ion-cyclotron unstable in runs RB, R4, RC and R3 at $t=0$, with $\gamma_m/\Omega_{ci}$ $\gtrsim$ $10^{-3}$. 
However, the growth rate rapidly drops as $T_\perp/T_\parallel$ decreases, and in fact the ion-cyclotron instability is not disruptive and it has no quantifiable effect on the dynamics of the Alfv\'en wave \citep{klein2015impact,kunz2018turb}. The growth rate from NHDS drops to nearly zero values (see also the black dots in Fig.~\ref{fig:parspace1}), before picking up again the firehose instability at much later times. 

{ In Figure \ref{fig:anisotropy}, the evolution of $T_\perp/T_\parallel$ for each run shows an initial rapid decrease, caused by an increase in $T_{\parallel}$ associated with beam formation. However, this decrease is not sufficient to trigger the firehose instability, which also requires a large plasma beta ($\beta\gtrsim1$) \citep{matteini_ion_2012}. As the system evolves, the expansion drives the anisotropy well below unity while $\beta_\parallel$ continues to increase, until the system eventually becomes firehose unstable at much later times, on average at around $t = 10000\;\Omega_{ci}^{-1}$. This transition can be inferred from the rapid increase in $T_\perp/T_\parallel$ in Fig.~\ref{fig:anisotropy} and is also indicated in Fig.~\ref{fig:crosshl1}).} 

Between the early times and the onset of the firehose, ALPS provides a more accurate evaluation of the unstable modes than NHDS, as the VDF evolves from a bi-Maxwellian at $t=0$ to a core-beam system that includes an additional population of trapped particles generated by parametric decay. 
The dotted and starred colored lines reported in Figure \ref{fig:growthrates} represent the maximum growth rate obtained from ALPS for the unstable fast and Alfv\'en mode, respectively, for three selected cases. 

By inspection of Figure \ref{fig:growthrates}, one can see that when the beam is formed, the Alfv\'en mode has a large growth rate (larger than the expansion rate, the first starred data points in the plot), suggesting that kinetic instabilities are responsible for slowing down the beam as soon as it forms (see  Fig.~\ref{fig:temp_drift}). More specifically, results from ALPS show that both the Alfv\'en/ion cyclotron and the fast magnetosonic branches are unstable, with the Alfv\'en mode however displaying a larger growth rate. This result is in agreement with predictions of core-beam drift instabilities in the presence of core anisotropic temperatures. In fact, although the total anisotropy of the proton VDF is $T_\perp/T_{\parallel}<1$, the  core and/or beam can exhibit $(T_\perp/T_{\parallel})_{b,c}>1$. The presence of such anisotropies tends to reduce the growth rate of the fast magnetosonic mode (otherwise driven by the relative drift), whereas   the presence of a mono-directional beam at relatively small core anisotropy allows only the forward ion-cyclotron unstable mode to grow  \citep{araneda2002proton}. Remarkably, it has been shown via numerical simulations that both ion-cyclotron and magnetosonic unstable branches slow down the relative drift \citep{araneda2002proton}.


By comparing Figure \ref{fig:growthrates} with Figure \ref{fig:temp_drift}, we note that RB shows a long-term trend of $v_d/v_a$ that is in agreement with expansion-driven evolution, while R4 and R2 display a slower than linear increase of the relative drift. 
Accordingly, R4 and R2 have a growth rate that is larger or comparable to the expansion rate, $\gamma/\Omega_{ci}\gtrsim10^{-4}$, while RB has a growth rate  $\gamma/\Omega_{ci}\lesssim10^{-4}$. 
For run RB, however, the Alfv\'en mode's growth rate increases with radial distance becoming eventually larger than the expansion rate. At that point, the Alfv\'en mode instability is comparable to the fast mode (firehose) and the effect of the two instabilities on the beam is effectively not discernible. 

At the onset of the parallel fire-hose instability, shown also by the increase in perpendicular heating (Fig. \ref{fig:invariants}), there is  rebounding below the $\gamma_m$ = $10^{-3}$ threshold (red dashed line in Fig.~\ref{fig:parspace1}) similar to other work \citep{hellinger_solar_2006}. 
In that final stage, $\gamma_m$ for each run eventually saturates at $\gamma_m/\Omega_{ci}\gtrsim10^{-2}$ as a sort of equilibrium between firehose-induced heating and expansion-driven cooling is achieved. 
Ultimately, it is the firehose instability to scatter the beam in phase space and to modify the global structure of the VDF, thereby strongly affecting plasma heating. 

\section{Comparison with Helios and Ulysses Data}
\label{data_comparison}

In Figure \ref{fig:parspace1},  we compare the evolution of the simulations with solar wind data in the $(\beta_{\parallel}, T_\perp/T_\parallel)$ parameter space. 
Each run's evolution is also compared with the CGL prediction (dashed line) and the empirical fit made by fitting Helios data (dot-dashed line) \citep{matteini_signatures_2013}. 
Due to expansion, the background magnetic field  decreases as $(R/R_0)^{-2}$ as well as the total density due to the continual volume increase. 
Then the invariants (eqn. \ref{eqn:cglpars}) give the CGL prediction, $T_{\perp}/T_{\parallel}\propto\beta_{\parallel}^{-1}$. 
When unaffected by instabilities, the slopes of each run follow this prediction very well. 
As discussed in section \ref{stage2}, parallel temperatures remain mostly constant during this quasi-stable state, whereas expansion drives the continual decrease of perpendicular temperatures to match the decreasing temperature anisotropy (Fig.~\ref{fig:invariants}). 
As a consequence the simulations do not reproduce the empirical relation $T_{\perp}/T_{\parallel}$ $\approx$ $\beta_{\parallel}^{-0.55}$ \citep{matteini_signatures_2013}, obtained from Helios and Ulysses data between 0.3 and 1 AU. This is due to missing perpendicular dynamics and turbulent cascade and dissipation that are not captured in 1D simulations. 

\begin{figure}[h]
\centering
\includegraphics[width=\linewidth]{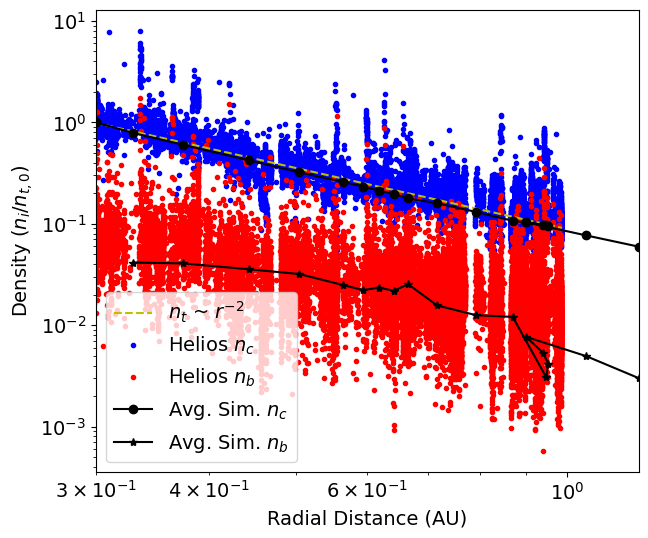}
\caption{The averaged simulation profiles of core and beam density, normalized to the total density at 0.3 AU, $25\;cm^{-3}$. 
The red/blue dots represent Helios beam/core density data respectively.}
\label{fig:partdens}
\end{figure}

While our simulations do not reproduce the observed perpendicular heating,  comparisons reported in 
Figure \ref{fig:partdens} and Figure \ref{fig:parspace1} of core and beam properties show good agreement between simulations and data. 

Figure \ref{fig:partdens} displays the ensemble-averaged core and beam densities from all simulation runs (black dots), overlaid with Helios observations shown in blue and red, respectively. 
All densities are normalized to the total proton density, $25\;cm^{-3}$, at $R=0.3$ AU for both the Helios and simulation datasets, displaying quite a good agreement during solar wind expansion with consistent relative densities throughout this period.  

\begin{figure}[h]
\centering
\includegraphics[width=1\linewidth]{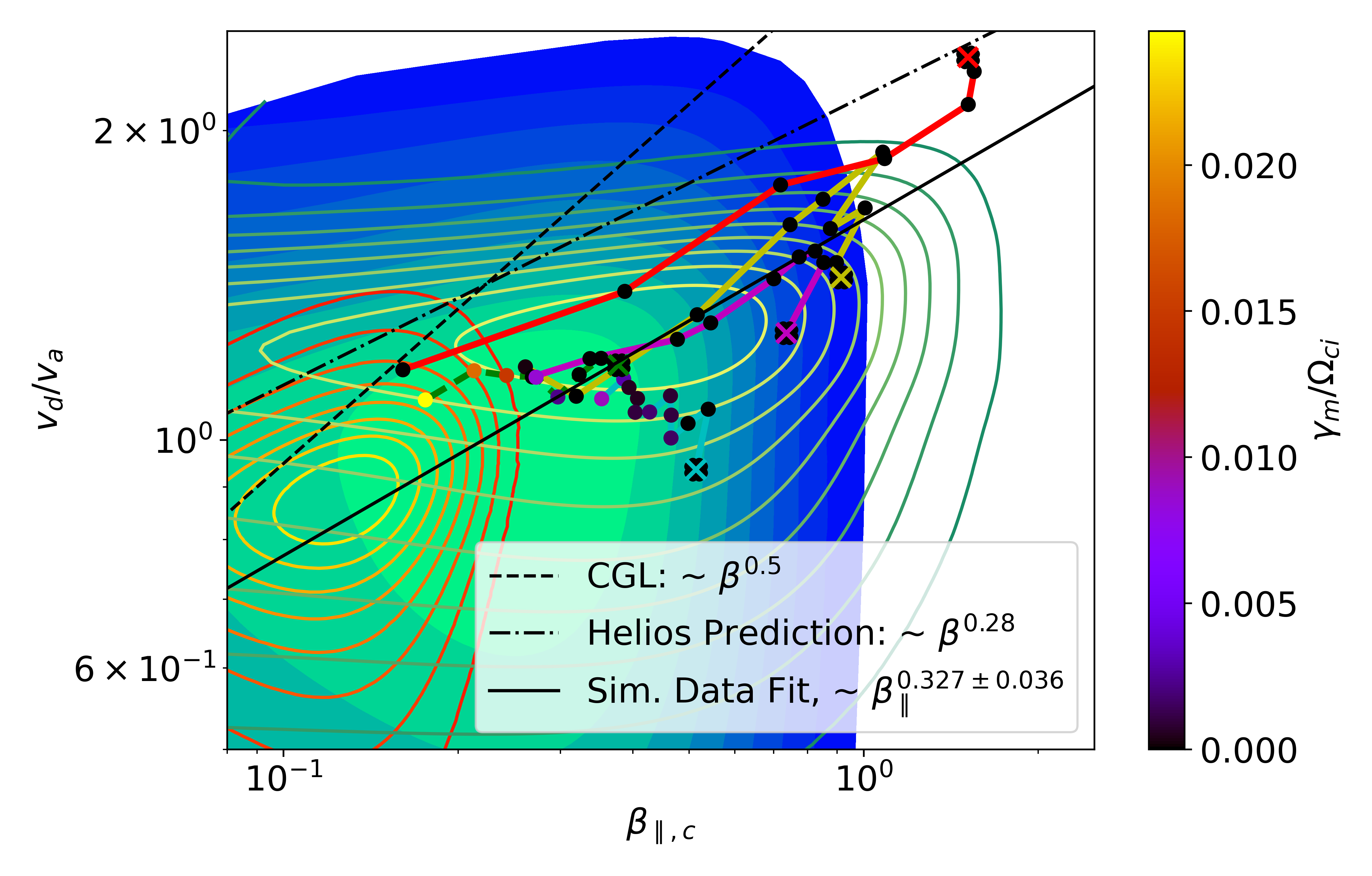}
\caption{Parameter space of normalized drift vs core parallel beta with the background taken from Ulysses (green contour) and Helios (red and blue for 0.3 and 1 AU respectively) fast wind data. 
The paths are color coded by the maximum growth rate at that point in time. 
The x marks the time at which, after the VDF collapse, the beam was indiscernible from the total VDF.}
\label{fig:parspace2}
\end{figure}

Figure \ref{fig:parspace2} reports simulation data points of core parallel plasma beta  $\beta_{\parallel,c}$, and normalized drift $v_d/v_a$ from each run, colored by maximum $\gamma_m$ from NHDS results at each point. 
Contours represent Helios and Ulysses fast solar wind data with contour colors matching the colors from Figure \ref{fig:parspace1}. 
Core parallel beta and normalized drift is calculated from in-situ measurements of core proton temperature, magnetic field magnitude and proton velocities. 
CGL evolution together with expansion  predict $v_d/v_a$ $\approx$ $\beta_{\parallel,c}^{0.5}$, represented by the overlaid dashed line. 
However, a best fit on the ensemble of simulations yields the scaling   $\beta_{\parallel,c}^{0.327}$, which is close to the Helios data extrapolation  of $\beta_{\parallel,c}^{0.28}$, shown by the dot-dashed line \citep{tu_dependence_2004}. 
Since $\beta_{\parallel,c}$ evolves close to the CGL prediction (we find $\beta_{\parallel,c}\propto (R/R_0)^{1.7\pm 0.1}$ on average over the runs), the deviation from CGL trends observed in this parameter space is caused by the sub-linear trend with $R$ of the normalized drift (Fig.~\ref{fig:temp_drift}). 
The further the normalized drift is from a linear evolution in radial distance, the more drift unstable the system will be as it evolves in parameter space.

\section{Discussion and summary}
\label{discussion}

By using hybrid simulations, we have studied the formation and evolution of a proton beam driven by a large amplitude Alfv\'en wave under solar wind expansion, by considering a range of initial conditions to account for different solar wind plasma states. 

We have shown that a field-aligned proton beam forms at the steepened edge of the Alfv\'en wave drifting at the local Alfv\'en speed with respect to the core population. 
The initial steepening leading to the beam acceleration is followed by the generation of ion acoustic modes by parametric decay.  
This result is in agreement with prior work that however did not consider expansion \citep{machida_simulation_1987, buti2000hybrid, gonzalez2021proton}, and is consistent across the different initial conditions considered. 

With expansion, the core-beam system slowly evolves with radial distance, but persists at around the local Alfv\'en speed in all runs until the onset of the fire-hose instability. 
The normalized drift speed $v_d/v_a$ increases with $R$, at a rate that depends on the specific case considered (and with the exception of R4, which is the most unstable when the beam is formed in the first place, and also has the largest beam density owing to the higher $\beta$). 
Such increase is, on average over all simulations, slower than linear ($v_d/v_a\propto R^{0.61\pm0.06}$, Fig.~\ref{fig:temp_drift}), contrary to predictions based solely on expansion. 
By making use of ALPS on selected cases, we have shown that there exists an unstable Alfv\'en mode branch and magnetosonic branch, with the former being the dominant unstable mode, suggesting that kinetic instabilities cause the observed slow-down of the drift \citep{araneda2002proton}. 

The comparisons in Figure~\ref{fig:partdens} and \ref{fig:parspace2} reveal a good agreement in the relative core and beam densities between the simulations and Helios measurements, as well as the average trend of $v_d/v_a$ vs $\beta_{\parallel,c}$ ($v_d/v_a\propto\beta_{\parallel,c}^{0.327\pm0.036}$). 
Such quantitative agreements indicate that our simulations successfully capture the main properties of the core-beam dynamics in the solar wind and support the general assumption that the observed sub-linear evolution of $v_d/v_a$ in the solar wind is due to the core-beam instability.

Turbulence, wave-particle interactions  and kinetic instabilities are expected to contribute to heating or cooling the plasma.  
This work focuses on the field-aligned dynamics and the geometry chosen inhibits the perpendicular turbulent cascade. 
Consequently, {perpendicular heating mediated by the turbulent cascade is not captured}, explaining why the perpendicular adiabatic invariant $C_\perp$ is conserved after the initial transient stage, until the firehose instability sets in. 
On the other hand,  two competing mechanisms are at play in the field-aligned direction, with parametric decay leading to parallel heating \citep{gonzalez2023particle}, and wave-particle scattering reducing the core-beam drift, likely from kinetic instabilities \citep{araneda2002proton}. 
In our simulations, parallel heating by parametric decay can be clearly observed in the transient sharp increase in $C_\parallel$ at times $t<1000$, {before reaching a nearly constant value of $C_\parallel\approx 1.4\times 10^{6}\, KnT^{2}cm^{6}$ ($\approx 2\times10^{-47}\,JT^2 m^6$), found by taking the average of $C_{\parallel}$ (fig.~\ref{fig:invariants}, lower panel) across all runs and then converting from simulation units.} 
Analysis of Helios data have reported that parallel heating rate is smaller than the perpendicular one, and that it transitions from heating to cooling \citep{hellinger2011heating}. 
However, it is possible that such estimates are within uncertainties as the total $C_\parallel$ is observed to slightly oscillate with radial distance, but it remains nearly constant in the range $0.2\lesssim R\lesssim1$ \citep{zaslavsky2023evaluation}. 
Based on our results, we suggest that the observation of a nearly conserved  $C_\parallel$ can be interpreted as the combined effect of parametric decay or other mechanisms mediated by compressible fluctuations, and particle scattering due to kinetic instabilities \citep{goldstein2000observed, matteini_signatures_2013}.

In conclusion, despite its simplified  geometry, our model reproduces  observed features of the solar wind proton VDF when considering an ensemble of possible  plasma states.
The one-dimensional simulations performed here are however limited, in that fluctuations with perpendicular gradients have not been considered. {As such, unstable oblique modes are not captured, although the general scenario of expansion-driven kinetic instabilities slowing down the beam would likely remain unaffected. However}, processes involving oblique modes that are known to modify and weaken parametric decay as well as mediate the perpendicular turbulent cascade, are not captured. 
These effects will be investigated in a followup study where we will consider a 3D spectrum to investigate parallel and perpendicular wave dynamics and their effects on particles.

\begin{acknowledgements}
    {We acknowledge support from the NASA grants 80NSSC20K1275 and 80NSSC18K1211 and the NSF CAREER award 2141564.
    KGK was supported in part by NASA grant 80NSSC24K0724.
    The authors also acknowledge the Texas Advanced Computing Center (TACC) at The University of Texas at Austin for providing HPC resources that have contributed to the research results reported within this paper. URL: \url{http://www.tacc.utexas.edu}. We are grateful to M. Mihailovic for providing access to the fitted Helios data used in this work and to D. Verscharen for his help with NHDS.}
\end{acknowledgements}

\bibliography{reference}{}
\bibliographystyle{aasjournalv7}

\end{document}